\documentclass[a4paper,12pt,onecolumn,final]{panl}
\usepackage{graphicx,amsmath,amssymb}
\usepackage{cite}
\usepackage{xcolor}
\def\rev#1{\textcolor{red}{#1}}
\def\be{\begin{eqnarray}}
\def\ee{\end{eqnarray}}
\def\om{\omega}
\newcommand{\lsim}{\stackrel{\scriptstyle <}{\phantom{}_{\sim}}}
\newcommand{\gsim}{\stackrel{\scriptstyle >}{\phantom{}_{\sim}}}

\originalTeX

\begin{document}\title{  Charged scalar  bosons under rotation and acceleration}
\maketitle
\authors{M. Bordag$^{\,a,b}$, D.\,N.\,Voskresensky$^{\,a,c}$}
\from{$^{a}$BLTP, Joint Institute for Nuclear Research, RU-141980 Dubna, Russia}
\from{$^{b}$Institute for Theoretical Physics, University Leipzig,
		IPF 231101, D-04081 Leipzig, Germany}
\from{$^{c}$National Research Nuclear University ``MEPhI'', 115409 Moscow, Russia}

\begin{abstract}  Creation of  charged spinless bosons  from vacuum in the rigidly rotating frame is studied  in presence of the external static electromagnetic fields, being formed either in the resting, or in the general rotating or in the local-flat  frames. It is shown that the description remains the same  in the resting and the local-flat frames.  More specifically, the case of a solenoid (magnetic flux tube) embedded into the rotating empty  cylinder (rotation frame) is studied in presence or absence of a static square electric potential well $eA_0$.   A supervortex  of a spinless-boson field can be created from vacuum when the rotation frequency exceeds a critical value $\Omega_c$. It is shown that $\Omega_c$ is smaller  provided the solenoid  rests in the resting frame.  Then the case is considered when the rotating frame additionally  moves with acceleration $\vec{w}(\vec{r})\neq 0$. A specific case  $w(r)=G/r$ for $G=const$ is treated explicitly. It is shown that for $G<0$ the charged spinless-boson vortex  can be created in rapidly rotating system even in the limit when  the magnetic  and electric fields tend to zero.
\end{abstract}
\vspace*{6pt}

 \maketitle
\section{Introduction}

Boson vacuum can become unstable in strong external  fields and in the dense baryon matter, cf. \cite{Migdal:1971cu,
Grib1980,
MSTV90
}  and references therein.   Recently, question on instability of the vacuum of charged pions at relativistic rotation  under action of a  strong external uniform magnetic field was  studied in  \cite{Zahed,Guo,Voskresensky:2023znr,Voskresensky:2024ivv,Voskresensky:2024vfx,
Bordag:2025ets,Bordag:2025xwp,BV2025}.

The vorticity with  rotation frequency  $\Omega\simeq
(9\pm 1) \cdot 10^{21}$ Hz $\simeq 0.05m_\pi$, $m_\pi\simeq 140$\,MeV is the pion mass, is formed in peripheral heavy-ion collisions of Au $+$ Au  at $\sqrt{s} = 200$ GeV, see \cite{Adamczyk}, which corresponds to angular momenta $l\lsim 10^6$, $\hbar=c=1$.
In addition to rotation, also strong magnetic fields $H\lsim 10^{18}$G are expected to occur in heavy-ion collisions and in compact stars, cf. \cite{Voskresensky:1980nk}.
Specific properties  of the rotating quark-gluon plasma were found in lattice calculations, cf. \cite{Braguta:2023tqz}.
 Effects of accelerated reference frame at initial stage of  heavy-ion collisions were discussed in \cite{Prokhorov:2025vak}. Finally, as   is well known   vortex structures are formed in rotating nonrelativistic superfluids and superconductors, cf. \cite{Tilly-Tilly}.

The given Letter is organized as follows.
Sec. \ref{chir} describes behavior of the vacuum of charged spinless bosons  in the presence of electromagnetic fields $A_\mu(x_\nu)$ formed either in the resting frame, or in the local-flat frame, or in  the general rotation frame, see  \cite{BV2025}.  Sec. \ref{solenoid} studies  a possibility  of the occurrence of  instability of the spinless-boson vacuum  in presence of the  solenoid (magnetic flux tube)  in presence or absence of a square electric potential well  embedded  in the empty rapidly rotating cylinder (associated  with the rotation frame). We study differences of cases when the external fields have such a form either in the resting or in the rotation frames.
Simplifying consideration we focus on the single-particle problem, i.e. on the  case of $\Omega\leq \Omega_c$, where $\Omega_c$ is a critical rotation frequency.
Sec. \ref{acceleration} discusses influence of a local acceleration  of the rotation frame.

\section{Charged spinless bosons in  rotation frame
}\label{chir}
We  study behavior of charged spinless bosons  in a rigidly rotating cylindrical system of coordinates $(r,\theta, z)$ at the constant rotation frequency $\vec{\Omega}\parallel z$,    $r=\sqrt{x^2+y^2}$,  $\nabla =(\partial_r, \partial_\theta
/r, \partial_z)$. We use standard coordinate transformation   \cite{LL2} between the laboratory (resting) frame $(t_R,\vec{r}_{R})$  and the general rotation frame  $(t,\vec{r})$:
\begin{eqnarray}&t_R=t, \,x_R=x\cos (\Omega t)-y\sin (\Omega t), y_R=x\sin (\Omega t) +y\cos (\Omega t),\,z_R=z.\label{Galilei}\end{eqnarray}
Employing  that
$(ds_R)^2=\delta_{\mu\nu}dx^{\mu}_R dx^{\nu}_R=(ds)^2=g_{\mu\nu}dx^\mu dx^\nu\,,$
where  $\delta_{\mu\nu}=\mbox{diag}(1,-1,-1,-1)$, one gets
  the interval in the general rotation frame,
  \be (ds)^2=(1-\Omega^2 r^2)(dt)^2 +2\Omega y dx dt  -2\Omega x dy dt -(dr_3)^2\,,\nonumber
  \ee
 $r_3= \sqrt{r^2 +z^2}$. Greek indices run values $0,1,2,3$.   The rigidly rotating system must be finite to fulfil  the causality condition, $\Omega r<1$. Tensors $g^{\mu\nu}$ and $g_{\mu\nu}$ in the rotation frame under a local acceleration  are presented in
 Appendix.

 Let us first  focus on case of the rotation  at absence of acceleration,  $\vec{w}=0$.
We introduce transformation matrices \cite{LL2,BV2025}, $T_\mu^{\,\,\nu}=\frac{\partial x_R^\nu}{\partial x^\mu}$, and $\widetilde{T}_\mu^{\,\,\nu}=\frac{\partial x^\nu}{\partial x_R^\mu}$,
\begin{eqnarray}
T_{\mu}^{\,\,\nu}=
  \begin{pmatrix}  1 & -y_R \Omega & x_R\Omega&0 \\
   0 & \cos (\Omega t) & \sin (\Omega t)&0 \\
   0&-\sin (\Omega t) & \cos (\Omega t) &0 \\
   0 & 0 & 0&-1 \\
\end{pmatrix},
\widetilde{T}_{\mu}^{\,\,\nu}=
  \begin{pmatrix}  1 & y \Omega & -x\Omega&0 \\
   0 & \cos (\Omega t) & -\sin (\Omega t)&0 \\
   0&\sin (\Omega t) & \cos (\Omega t) &0 \\
   0 & 0 & 0&-1 \\
\end{pmatrix}.\nonumber
\end{eqnarray}
 The  co-variant 4-vector field $A_\nu (x^\delta)$ is transformed from the general rotation frame to the resting frame and backward as
\be
A_\mu (x_R^\delta)=T_{\mu}^{\,\,\nu} (x_R^\delta) A^R_\nu (x_R^\delta)\,,\,\,\,A^R_\mu (x^\delta)=\widetilde{T}_{\mu}^{\,\,\nu}(x^\delta) A_\nu (x^\delta)
.\label{Ttrans}
\ee
 Applying the $\widetilde{T}_{\mu}^{\,\,\nu}$ operator to $A_\nu (x^\delta)= (A_0(x^\delta), -A^x(x^\delta), -A^y(x^\delta),0)$  we find
 \be A_\mu^{\rm R}(x^\delta)=(A_0- y\Omega A^x+ x\Omega A^y, -A^x_R, -A^y_R, 0)\,,\label{Arest}
\ee
being expressed in variables $x^\delta$ of the general rotation frame. We used that $\vec{A}$ is transformed following  (\ref{Galilei}). Only zero component of the 4-vector field is changed. The spatial components $A_i$, $i=x,y,z$, remain the same as in rotation frame.
We will consider  fields of the form
\be\vec{A}=\vec{e}^{\,\,\theta} \tilde{a} (r)/r\label{ansatz}
\ee  in cylindrical coordinates, where $\vec{e}^{\,\,\theta} =(-y/r, x/r,0)$, $\mbox{div} \vec{A}=0$.
Thereby Eq. (\ref{Arest}) can be rewritten as
 \be A_\mu^{\rm R}(x^\delta)=(A_0(x^\delta)-\Omega \tilde{a}(r),-A^x_R(x^\delta),-A^y_R(x^\delta),0).\label{Arest1}
\ee
First relation (\ref{Ttrans}) yields
\be A_\mu (x^\delta)=(A_0^R(x_R^\delta(x^\delta))+\Omega  \tilde{a}(r),-A^x,-A^y,0).
\ee

The nonlinear Klein-Gordon equation  for the complex spinless-boson field interacting with the 4-vector gauge field $A^\mu$ in the rotation frame renders
 \begin{eqnarray}
 \frac{1}{\sqrt{-g}}D_\mu(\sqrt{-g}g^{\mu\nu}D_\nu \phi)+m^2\phi +\lambda|\phi|^2\phi=0\,,\label{KGFrot}
\end{eqnarray}
$m>0$ is the mass of the excitation of the spinless-boson field, e.g. $m\simeq 140$MeV for the pion, and $\lambda \geq 0$ is the coupling of the self-interaction.
Further to be specific, as $A_\mu$, we consider electromagnetic field, then $D_\mu=\partial_\mu +ieA_\mu (x^\delta)$, $e$ is the charge of the electron, $\sqrt{-g}=\mbox{det}\{\partial x^\mu/\partial x^\nu_R\}$. In the given case $\sqrt{-g}=1$.
Employing Eq. (\ref{gmunuac}) of Appendix for $\vec{w}=0$ we arrive at
 \begin{eqnarray}
&[(\partial_0 +ie A_0+\Omega y (\partial_x-ie A^x) -\Omega x (\partial_y-ie A^y)]^2\phi\label{Arotfr}\\
&-(\partial_x -ieA^x)^2\phi-(\partial_y -ieA^y)^2\phi +m^2\phi^2 +\lambda|\phi|^2\phi=0\,.\nonumber
\end{eqnarray}
It can be rewritten in terms of $A_\mu^R(x^\delta)$ as
 \begin{eqnarray}
&[\partial_0 +ie A_0^R (x^\delta) +\Omega y \partial_x -\Omega x \partial_y]^2\phi-(\partial_x -ieA^x_R(x^\delta))^2\phi \nonumber\\
&-(\partial_y -ieA^y_R(x^\delta))^2\phi +m^2\phi +\lambda|\phi|^2\phi=0\,.\label{KGrest}
\end{eqnarray}
Thus to describe $\phi (x^\delta)$ in rotation frame, in case if electromagnetic potential has the given  form in the resting frame, e.g. $e A_0^R=-V_0=const$ and $\vec{A}_R= (-Hx, Hy,0)$ with $H=const$  (see case (i) below),  we should employ  Eq. (\ref{KGrest}), whereas, if $e A_0=-V_0=const$ and $\vec{A}= (-Hx, Hy,0)$ with $H=const$ in the rotation frame (see case (ii) below), we should use Eq. (\ref{Arotfr}).

Another possibility is to describe rotation employing the transition from the general rotation frame to the local-flat frame, which can be performed using the tetrads \cite{LL2}.  The tetrad is determined as $e^{\hat{\alpha}}_{\,\,\mu}=\partial x^{\hat{\alpha}}/\partial x^\mu$, $e_{\hat{\alpha}}^{\,\,\,\mu}=\partial x^\mu/\partial x^{\hat{\alpha}}$, $g^{\mu\nu}=\eta^{\hat{\alpha}\hat{\beta}}e_{\hat{\alpha}}^{\,\,\,\mu}e_{\hat{\beta}}^{\,\,\,\nu}$,
$\eta_{\hat{\alpha}\hat{\beta}}=\{+,-,-,-,\}$,
 \begin{eqnarray}
 &e^{\hat{\alpha}}_{\,\,\mu}=\delta^{\hat{\alpha}}_{\,\,\mu}+\delta^0_{\,\,\mu}
 \delta^{\hat{\alpha}}_{\,\,i}  v^i\,,\quad
 e_{\hat{\alpha}}^{\,\,\,\mu}=\delta_{\hat{\alpha}}^{\,\,\,\mu}-\delta_{\hat{\alpha}}^{\,\,\,0} \delta_i^{\,\,\mu} v^i\,,
 \label{tetrades}
 \end{eqnarray}
 $\vec{v}=(0, v^1,v^2,v^3)=[\vec{\Omega}\times \vec{r}_3]$. Latin index $i=1,2,3$, greek tetrad and Lorentz indices are $\hat{\alpha},\mu =0,1,2,3$;
  \be\partial_{\hat{\alpha}} +ieA_{\hat{\alpha}}=e_{\hat{\alpha}}^{\,\,\mu}(\partial_\mu+ieA_\mu)\,,
  \ee
  $\partial_{\hat{0}}=\partial_t +y\Omega\partial_x -x\Omega\partial_y=\partial_t -i\Omega\hat{l}_z=
\partial_t -\Omega \partial_\theta\,,$ $e_{\hat{i}} =\partial_i$,
   \be A_{\hat{\alpha}}(x^\delta)=(A_0-\Omega y A^x+ x\Omega A^y, -A^x, -A^y, 0)\,,\label{flA}
\ee
$\Omega\hat{l}_z=\vec{\Omega}[\vec{r}_3\times \hat{\vec{p}}]$, $\hat{\vec{p}}=\nabla/i$.
 Thus the nonlinear Klein-Gordon equation in the local-flat frame,
 \be [(\partial_{\hat{\alpha}}+i eA_{\hat{\alpha}})^2 +m^2]\phi+\lambda|\phi|^2\phi =0\,,\label{lfeq}
 \ee
can be rewritten as
\begin{eqnarray} &(\partial_t+ieA_{\hat{0}}(x^\delta) +y\Omega\partial_x -x\Omega\partial_y)^2\phi -(\partial_x -ieA^{\hat{x}}(x^\delta))\phi \nonumber\\ &-(\partial_y -ieA^{\hat{y}}(x^\delta))\phi +m^2\phi+\lambda|\phi|^2\phi=0\,.\label{KGFlocal}
 \end{eqnarray}
Variables are expressed in terms of the general rotation frame.  $A_{\hat{\mu}}$ can be rewritten in terms of the fields determined in the rotation frame employing Eq. (\ref{flA}).
We see that Eqs. (\ref{KGrest}) and (\ref{KGFlocal}) coincide provided $A^R_\mu=A_{\hat{\mu}}$.
{\em This is one of the key messages of this Section.} Thereby we further will not distinguish whether we deal with the resting or the local-flat frame.

Let us focus  on  the case of the static scalar (or pseudoscalar)  field, $\phi\propto e^{-i\epsilon t}$, $\epsilon =const$.
 The  Lagrangian density of the charged stationary spinless-boson  field in the local-flat frame is
\begin{eqnarray}
&{\cal{L}}_b={|\widehat{\widetilde{\epsilon}}\phi|^2 }-{|(\partial_i+ieA_{\hat{i}})\phi|^2}-{m^{2} |\phi|^2}-\frac{\lambda |\phi|^4}{2}\,,\label{murep}\\
&\widehat{\widetilde{\epsilon}}=\epsilon  +iy\Omega\partial_x-ix\Omega\partial_y-V(r)\,,\nonumber
\end{eqnarray}
$eA_{\hat{0}}=V(r)$.  Here variables are presented in the general rotation frame, whereas the (static) electromagnetic field  is determined in the local-flat frame. 
Variation of (\ref{murep})
yields equation of motion (\ref{KGFlocal}), now in stationary case. In this case the energy density of the boson system is as follows
\begin{eqnarray}
&E_b=\epsilon n_b  -{\cal{L}}_b\,,\quad n_b=\partial {\cal{L}}_b/\partial \epsilon= \phi^*\widehat{\widetilde{\epsilon}}\phi+c.c.\,,\label{EnerPhipi}
\end{eqnarray}
where $c.c.$ means complex conjugation.
 For $\lambda =0$ on the solutions of the equation of motion we have ${\cal{L}}_b=0$ and energy is equal to
 \be{\cal{E}}_b=\epsilon N_b= \epsilon\int d^3r n_b\,,\label{enbo}
  \ee
  i.e., it is the energy of the given level $\epsilon$ times number of particles occupying it.
  The angular momentum $\vec{J}=\vec{J}_b+\vec{J}_{el}$ contains   contributions of the charged boson field
\begin{eqnarray}&{\vec{J}}_{b}=\int d^3 X [ {\vec{r}}_3\times{\vec{P}}_b]\,\,,\quad
\vec{P}_b=-\frac{\partial{\cal{L}}_{b}}{\partial\partial_t \phi}{{\nabla}} \phi
 -\frac{\partial{\cal{L}}_{b}}{\partial\partial_t{\phi}^*}{{\nabla}}\phi^*
,\,\,\label{pphi}\end{eqnarray}
 and the electromagnetic field
\be
	\vec{J}_{el} = 2\pi\int_0^R dr\,r\ \vec{r}_3 \times (\vec{E}\times\vec{h}),\label{Jem}
\ee
where $\vec{E}=-\partial_0 \vec{A}-\nabla A_0$, $\vec{h}=\nabla\times \vec{A}$.
For $\phi\propto e^{il\theta-i\epsilon t+ip_z z}$, see Eq. (\ref{Phifieldformpi}) below, we obtain  $J_z^b=lN_b$, and for $A_0=A_0(r)$ and the vector potential of the form (\ref{ansatz}),
Eq. (\ref{Jem}) simplifies to
\be J_z^{el} = 2\pi\int_0^R dr\,r\ \tilde{a}'(r)A_0'(r).
\ee
It indicates that in configurations, where $A_0'(r)=0$ or $\tilde{a}'(r)=0$, the electromagnetic field does not contribute to the angular momentum.

  In the single-particle problem ($\lambda =0$) the  ground state boson energy level $\epsilon=\epsilon_{gr}$  decreases, e.g.,  with deepening of the electrical potential well, $V=-V_0$, reaching zero for $V_0>m$, and $-m$ for $V_0>2m$ in case of a broad potential well with the size  $R\gg 1/m$, even for $\Omega =0$, cf. \cite{Migdal:1971cu,MSTV90}. In  case  of the rotating system of our interest,
$\epsilon_{gr}$ still decreases with increasing  $\Omega$.
The vacuum becomes unstable when $\epsilon_{gr}$
reaches $-m$. Then  pairs of bosons can be produced in the tunneling process from the lower to the upper continuum.
In presence of matter (in our case in presence of the cylinder) instability may appear when $\epsilon_{gr}$ reaches zero, cf. \cite{Voskresensky:2023znr,Voskresensky:2024vfx,BV2025}.  In given Letter we assume that  $\Omega$ is fixed by  external conditions. The produced angular momentum and charge
in this case are compensated by the rotating cylinder. After instability appeared,  one should take into account $\lambda \neq 0$ and solve the problem for the classical field $\phi$. The energy (at condition $\Omega =const$) has minimum for the static field, so in this classical field  problem we put $\epsilon =0$, cf. \cite{Guo,Voskresensky:2023znr,Voskresensky:2024vfx}. For $\lambda>0$ employing (\ref{EnerPhipi}) for the static fields (for $\epsilon =0$),  (\ref{KGrest}), and using integration by parts one gets
${\cal{E}}_b=-\int d^3r \frac{\lambda |\phi|^4}{2}\,.$
Below  we will focus on the single-particle problem.

\section{Case of flux tube embedded into rotating cylinder}\label{solenoid}

Let us deal with empty   cylinder of the radius $R$ and assume that in its resting frame  there is the external electric potential  $eA_0 =V=const<0$ for $r<R$. Cylinder rotates with angular velocity $\Omega<1/R$ to fulfill causality. Inside rotating cylinder it is embedded the $r$-symmetric flux tube (solenoid) of the transversal radius $R_s\leq R$. We consider two cases: (i) flux tube rests in resting frame  forming there constant external  magnetic field $\vec{H}=(0,0,H)$ at $r<R_s$  and zero for $r>R_s$, obeying Eq. (\ref{KGrest}) in variables of rotation frame, (ii) flux tube rests forming at $r<R_s$  constant external  magnetic field $\vec{H}=(0,0,H)$  in the  rotation frame where $A_0\neq const$, obeying Eq. (\ref{Arotfr}) in this frame.

Let us use the Dirichlet boundary condition on the cylinder,
\be  \phi (r=R)=0\,.\label{boundcond}
\ee

We will seek solution  in the form of the individual vortex:
\be \phi =\phi_{0}\chi (r)e^{il\theta-i\epsilon t+ip_z z}\,,\label{Phifieldformpi}
\ee
where $\phi_{0}$,   $\epsilon$ and $p_z$ are  real constants,  $\chi(r)$ is the real function and, being interested in description of the minimal energy configurations, we put  $p_z=0$, $l$ is integer number. Not loosing generality we further take $l$ nonnegative.

Equation of motion, see Eq.  (\ref{KGrest})  in case (i), and Eq. (\ref{Arotfr}) in case (ii), for the condensate field renders, cf. \cite{Guo,Voskresensky:2023znr,Voskresensky:2024ivv},
\begin{eqnarray}
[\widetilde{\epsilon}^2+\hat{K} -m^{2}]\chi
 -\lambda |\phi_0|^2\chi^3 =0\,,\,\,\,\widetilde{\epsilon}=\epsilon  +\Omega l-V(r)+\zeta\,,\label{pipi}
  \end{eqnarray}
 \begin{eqnarray}
 \hat{K}=\Delta_r  -l^2/r^2-eHl -(eH)^2r^2/4 \quad {\rm for}\quad r<R_s
\,,\label{Kin}
 \end{eqnarray}
  \begin{eqnarray}
 \hat{K}=\Delta_r  -(l- \delta_\Phi)^2/r^2\,,\quad \delta_\Phi=|eH| R_s^2/2\quad {\rm for}\quad R_s<r<R\,,\label{Kout}
 \end{eqnarray}
 where  $\Delta_r=\partial_r^2 +{\partial_r}/{r}$,
 we took into account condition of conservation of the magnetic flux  $\Phi=\oint \vec{A}d\vec{L}=\pi HR_s^2$.   In case (i): $ \zeta=0$, and in case (ii): $ \zeta=\Omega r eA^\theta (r)$, $A^\theta (r)=Hr/2$ for $r<R_s$ and $A^\theta (r)=HR_s^2/(2r)$ for $r>R_s$.

{\bf Resting solenoid  in resting frame for  $R_s\simeq R$, $\lambda =0$.}

This is case (i). Here,   we deal with uniform  external constant magnetic field $H$ for all $r<R_s\approx R$
in the resting frame. This problem was considered  for $\lambda =0$, $V_0=0$ in \cite{Zahed}   and  for $\lambda > 0$ in \cite{Guo},  and including $V(r)=-V_0\neq 0$ in \cite{Voskresensky:2024ivv,Voskresensky:2024vfx}.
The solution of the equation of motion (\ref{pipi}) renders,
 \begin{eqnarray}
 \chi (r)=r^{|l|}{e^{-|e{H}|r^2/4}} _{1}F^1(-a, |l|+1, |e{H}|r^2/2)\,,\label{hypergeom}
\end{eqnarray}
 $_{1}F^1$ is a confluent  Kummer hypergeometrical function,
 \be
 a=-\frac{1}{2}(|l|-l+1)+\frac{{\mu}^2-m^{2}}{2|eH|}\,,\label{a}\ee
where quantity $\mu$ having sense of a chemical potential is
\be{\mu}=\epsilon +\Omega l +V_0\,.\label{barmu}
\ee

In case of a  homogeneous magnetic field in the whole space (i.e. in the limit $R\to\infty$), the solutions ``$a$''   are  integer numbers, $a_n (l,\delta_\Phi)=n$, $n=0,1,\dots$, they describe the Landau levels and the Kummer function turns into the Laguerre polynomials. These solutions are degenerate with respect to the orbital momentum $l$ with a degeneracy factor  $N=eH R^2/2=\delta_\Phi\geq l$, being the number of states in the interval $L_z\Delta k_z/(2\pi)$,  $L_z$ is the (infinite) longitudinal size of the cylinder.
The energy  is minimal for  nonnegative $l$. Thereby we further  restrict  to $l\ge0$. In general, the  relation $N\geq l$ does not apply for a field defined in a restricted space, e.g.  in case of the boundary condition at $r=R$, see  \cite{BV2025}. Also in the latter case there is no solution at $a =0$. The integer number $a$ corresponding to the minimal energy with which the boundary condition is fulfilled is $a=1$, corresponding to $N=l+1$.

For $l\geq 0$, Eq. (\ref{a})  can be rewritten as
  \begin{eqnarray}
 \epsilon =-V_0-\Omega l+\sqrt{m^{2}+|eH|(1+2a(l))}\,,\label{dispH}
 \end{eqnarray}
where we retained the solution, which yields $\epsilon\to +m$, as it should be  at the switching off the interaction and rotation.
 For arbitrary $a$ and $l$ not necessarily restricted by the condition $N\geq l$ the critical frequency follows from (\ref{dispH}) when one puts $\epsilon =0$,
 \be
 \Omega^H_c (a)=(-V_0+\sqrt{m^2+|eH|(2a+1)})/l\,.\label{omcrHmin1}
  \ee

The energy of the ground state, see (\ref{enbo}),  is given by
\be
\epsilon_{gr}N=\epsilon_{gr} L_z 4\pi\sqrt{m^{*2}+|eH|(1+2a)}\phi_{0}^2\int_0^R r dr  \chi^2\,,\ee
with $\chi$ from (\ref{hypergeom}). For $\epsilon_{gr}<0$ satisfying Eq. (\ref{dispH}),  the constant $\phi_{0}$ is not limited in case $\lambda =0$, provided bosons can be produced in reactions on walls of the solenoid and the cylinder. Otherwise instability arises for $\epsilon_{gr}<-m$. Stability is recovered when one includes the self-interaction $\lambda >0$ or takes into account conservation of the electric charge, see \cite{MSTV90,Zahed,Guo,Voskresensky:2023znr,Voskresensky:2024ivv,Voskresensky:2024vfx,
Bordag:2025ets,Bordag:2025xwp,BV2025}. in case $\Omega\neq 0$.

{\bf{Resting solenoid   in rotation frame at $R_s\simeq R$,  $\lambda =0$.}}

This is case (ii). Here   we deal with uniform  external constant magnetic field $H$ for all $r<R_s\approx R$
in the rotation frame. Then
from (\ref{Arotfr}) we get
\be\widetilde{\epsilon}(r)={\mu}+eH\Omega r^2/2\,
 \ee
 with ${\mu}$ from Eq. (\ref{barmu}). Let us use approximation
\be \widetilde{\epsilon}^2(r)\simeq {\mu}^2+{\mu} eH\Omega r^2+O(r^4)\,,\label{r4appr}
\ee
dropping terms $\propto O(r^4)$, that will be justified below.
Solution of Eq. (\ref{pipi}) is
\begin{eqnarray}
 \chi (r)=r^{|l|}{e^{-|e\tilde{H}|r^2/4}} _{1}F^1(-b, |l|+1, |e\tilde{H}|r^2/2)\,,\label{hypergeom1}
\end{eqnarray}
where $\tilde{H}=\sqrt{e^2H^2+4{\mu}\Omega|eH|}$,
\be
b=\frac{l |eH|}{2 |e\tilde{H}|}+\frac{{\mu}^2-m^{2}}{2|e\tilde{H}|}-\frac{|l|+1}{2}\,.
\ee

Then instead of Eq. (\ref{dispH}) we  obtain relation
  \begin{eqnarray}
 {\mu}^2 =m^{2}+|e\tilde{H}| (2b+1)-(|eH|-|e\tilde{H}|)l/2.\label{dispHii}
 \end{eqnarray}

Values of our interest are $mR\gg 1$, $|eH|\lsim m^2$, $\Omega R<1$ (in heavy ion collisions $\Omega R<0.5$). At these conditions approximation (\ref{r4appr}) indeed holds for $|\epsilon|\lsim m$. The integer number $b$ corresponding to the minimal energy with which the boundary condition is fulfilled is $b=1$, corresponding to $N=l+1$.

 Comparing (\ref{dispHii}) and (\ref{dispH}) for $a=b=1$ shows that the ground state energy in the later case is always smaller.   Thus we come to the conclusion that, if the solenoid of the radius $R_s=R$ was initially rotated with the cylinder of approximately the same radius, after a while it would be energetically profitable to stop the former.

{\bf{Resting thin flux tube in resting frame, $\lambda =0$.}}

The typical length scale in
 Eq. (\ref{pipi}) for $r>R_s$  with $\hat{K}$ from (\ref{Kout})   is
\be \xi_{>}=1/\sqrt{{\mu}^2-m^2}\,,\label{chiout}\ee
where $\xi_{>}$ has the meaning of a coherence length characterizing the scale of the  change of the field $\phi$. For $R_s/\xi_{>}\ll 1$, being the condition that flux tube is thin, as the boundary condition we may use
$\chi (R_s)=0\,,$
or $\chi' (R_s)=0$, instead of doing accurate matching of  $\chi(r)$   and $\chi^\prime (r)$ at $r=R_s$ with the solution decreasing towards $r=0$, being valid for $r\leq R_s$.   Thus
for $R_s/\xi_{>}\ll 1$ the solution of Eq. (\ref{pipi}) with $\hat{K}$ from (\ref{Kout})  can be considered as approximately valid for all $0<r<R$ and  we do not need to find the solution $\chi(r)$ for $r<R_s$. This was also demonstrated in \cite{BV2025} by numerical solution of the problem. Thereby up to  terms $O(R_s/\xi_{>})$ description of the thin flux tube remains the same independently, if the flux tube rests in the resting or the rotation frame.

Then as the boundary condition at $r\to 0$ we may use either
 \be
 \chi (0)=0\label{boundnul}
 \ee
or
  \be
 \chi (0)=1\,.\label{boundnul1}
 \ee
 Solution   (\ref{pipi}) satisfying the boundary conditions (\ref{boundcond}) and (\ref{boundnul})  for $l\neq \delta_\Phi$ and (\ref{boundcond}), (\ref{boundnul1}) for $l=\delta_\Phi$ produces
\be\chi(r)=J_{|l-\delta_\Phi|}(r/\xi_{>}) \,,\label{chiout1}
\ee
and the spectrum follows from fulfilment of condition (\ref{boundcond}):
\be \epsilon_{n,l}=-\Omega l -V_0 +\sqrt{m^2 +j^2_{n,|l-\delta_\Phi| }/R^2}\,,\label{spectrumout}
\ee
where $j_{n,l}$ denotes the zero of the Bessel function.
The  criterion of applicability, $R_s/\xi_{>}\ll 1$, is rewritten as $R_s j_{n,|l-\delta_\Phi|}/R\ll 1$, since (\ref{spectrumout}) yields $1/\xi_{>}^2=j^2_{n,|l-\delta_\Phi|}/R^2$, i.e., at least for $|l-\delta_\Phi|\lsim 1$ and $mR\gg 1$, inequality $R_s\ll R$ is sufficient in order to use condition (\ref{boundnul}) or (\ref{boundnul1}).

The   energy $\epsilon_{1,l}$ is minimal for $l=\delta_\Phi$. It is
\be \epsilon_{1,\delta_\Phi}\simeq -\Omega \delta_\Phi-V_0+\sqrt{m^2 +j^2_{1,0}/R^2}\,,\label{spectrumoutmin}
\ee
and for $mR\gg 1$ of our interest we have $\epsilon_{1,\delta_\Phi}\simeq -\Omega \delta_\Phi-V_0+m$. The vacuum becomes to be unstable producing the spinless-boson-vortex field provided
\be\Omega>\Omega_c \simeq (m-V_0)/\delta_\Phi\,.\label{omcrlstar}
\ee
 The ground state  level reaches zero  even for $V_0=0$.

 Let us show that instability remains even for  $l\gg \delta_\Phi\gg mR\gg 1$ and $\Omega =(1-\delta)/R$ for $\delta\ll 1$. In this case
\be
j_{n,l}\simeq l+cl^{1/3}+...,\quad c\simeq 1.856\,,\label{jabram}
\ee
 cf. \cite{Gradshtein}. Then, from (\ref{spectrumout}) we find
\be \epsilon_{n,l}\simeq -V_0+l\delta /R +cl^{1/3}/R -\delta_\Phi/R +...
\ee
As we see, even for $V_0=0$ the ground state  level reaches zero and  instability resulting in creation of the complex spinless-boson field appears for
$0<\delta <(\delta_\Phi-cl^{1/3})/l\,.$
Thus instability remains at least for  $l$ satisfying the condition  $0<(\delta_\Phi-cl^{1/3})/l\ll 1$, i.e. for $ l\ll (\delta_\Phi/c)^3$.

Finally we note  that the magnetic field  for $r>R_s$ satisfies equation $h_z=(\mbox{curl} \vec{A})_z=\frac{1}{r}\frac{\partial (rA_\theta)}{\partial r}=0$, as  in the Aharonov-Bohm effect, cf. \cite{Voskresensky:2024vfx,BV2025}.

\section{Spinless-boson vacuum under action of  acceleration and rotation}\label{acceleration}

To describe the  behavior of the spinless boson in locally  accelerated (with acceleration $\vec{w}(r)$) and  rotated (with rotation frequency $\Omega$) frame we will employ the metric introduced  in Ref. \cite{Nelson1987}.
It is described in Appendix \ref{Ap}.
In case of the rigid rotation we associated the reference frame with the rotation of empty cylinder of radius $R<1/\Omega$.
We avoid discussion of subtleties connected with description of rigid arbitrary moving frames employing approach of \cite{Nelson1987}.  To introduce  the local acceleration of the reference frame in a physical system we assume that the cylinder is filled by very heavy particles,  and we deal with local acceleration of these  particles. Their microscopic description  is not of our interest here.  Our aim  will be only  to demonstrate possibility of new effects for a spinless-boson field feasibly occurring in presence of a local  acceleration or deceleration in addition to the rotation of the system.  For this we will focus on a specific case of acceleration or deceleration of the form
 \be\vec{w}=G\vec{r}/r^2\,,
  \ee
  at $G=const$.
Then employing the metric tensor (\ref{gmunuac}) we may rewrite Eq. (\ref{KGFrot}) as follows
\begin{eqnarray}
&(1+G)^{-2}(\partial_0-x\Omega \partial_y +y\Omega \partial_x)^2\phi -(\partial_x^2+\partial_y^2)\phi+m^2\phi +\lambda |\phi|^2\phi =0\,.
\end{eqnarray}
We for simplicity  put $A_\mu =0$.
For $\lambda =0$ this equation has solution
$\chi(r)=J_{l}(r/\xi_{>})$
satisfying the boundary conditions  (\ref{boundnul}) for $l\neq 0$ and (\ref{boundcond}), now with
\be \xi_{>}=1/[(\epsilon+\Omega l)^2(1+G)^{-2}-m^2]^{1/2}\,,\ee
for $\xi_{>}^2>0$, i.e. for $\Omega>\Omega_c$. The energy spectrum is given by
\be \epsilon_{n,l}=-\Omega l+(1+G)\sqrt{m^2+j^2_{n,l}/R^2}\,,\label{accep}\ee
and the instability arises for
\be\Omega>\Omega_c=(1+G)l^{-1}\sqrt{m^2+j^2_{n,l}/R^2}.\ee
At $l\gg mR\gg 1$ using expansion (\ref{jabram})
we find that  instability occurs  at $\Omega>\Omega_c\approx (1+G)/R$.
Thus even a weak deceleration, $G<0$, of the empty relativistically rotating cylinder in the $r$ direction may result in occurrence of the instability of the spinless-boson-vortex field, even for $V_0=0$.

Generalization to the case of presence of the flux tube resting in the resting frame for $R_s =R$ and for  $V=-V_0=const$ is straightforward. As the result, Eq. (\ref{dispH}) is transformed to
  \begin{eqnarray}
 \epsilon =-V_0-\Omega l+(1+G)\sqrt{m^{2}+|eH|(1+2a(l))}\,.\label{dispHac}
 \end{eqnarray}
Eq. (\ref{spectrumout}) applicable for the infinitely thin flux tube  is modified as
\be \epsilon_{n,l}=-\Omega l -V_0 +(1+G)\sqrt{m^2 +j^2_{n,l-\delta_\Phi }/R^2}\,.\label{spectrumoutac}
\ee

Note that in heavy-ion collisions we may be, indeed, deal with a decelerating and rapidly rotating nuclear fireball under the action of electromagnetic field, which effects  may stimulate formation of  a giant pion vortex. For  collision energies $\sim A$ GeV  the fireball size and velocity at the compression stage can be estimated as  $R_0\sim 5m^{-1}_\pi$ and $v\sim c$. The deceleration at the compression stage is $w\sim c^2/R_0$,
that yields   $-G\lsim 1$ for $w\sim G/R_0$. Then the prepared nuclear fireball expands up to the freeze-out stage when typically $R_{\rm f.o.}\lsim 10m^{-1}_\pi$ fm.  For the expanding ideal nonrelativistic classical gas with adiabatic index $5/3$   for $t>0$  one has $R(t)=\sqrt{u^2 t^2+R_0^2}$, at constant $u\lsim (0.5-0.7)c$, \cite{Bondorf}. The expansion velocity is $\vec{v}=\vec{r} \dot{R}/R$, and the acceleration is $\vec{w}=\dot{\vec{v}}$, the dot means time derivative.  We estimate  $v_r< 0.5 c$ on the expansion stage and obtain a typical deceleration to be   $-G\lsim 0.5$. With these findings all three expressions  (\ref{accep}), (\ref{dispHac}), (\ref{spectrumoutac}) demonstrate possibility of the formation of the pion vortex field in heavy-ion collisions.

\section{Conclusion}

This paper continues  study \cite{Zahed,Guo,Voskresensky:2023znr,Voskresensky:2024ivv,Voskresensky:2024vfx, BV2025} of the possibility of the creation of the spinless bosons, e.g., charged pions from the vacuum at the rotation in presence  of static external electromagnetic fields of a special form. More specifically we study behavior of spinless bosons  in presence of a flux tube (solenoid) placed in the center of the rotating empty cylinder.

In Sec. \ref{chir}
it is explicitly demonstrated that solutions of the Klein-Gordon equation formulated in the variables of the general rotating frame at the rotation with the constant angular velocity, $\vec{\Omega}\parallel z$, coincide provided the electromagnetic field has the same form in the resting and the local-flat reference frames.
This question is studied to avoid ambiguities met in the literature.

Section \ref{solenoid} deals with empty rotating   cylinder of the radius $R<1/\Omega$ to fulfill causality in presence of the external electric potential, $eA_0=-V_0\leq 0$ for all $r<R$ with $V_0=const<0$.  Two cases are considered: (i) the flux tube of radius  $r<R_s$ forming  the uniform constant external  magnetic field $\vec{H}=(0,0,H)$ for $r<R_s$ and $\vec{H}=0$ for $r>R_s$, rests in the resting frame (and in the local-flat frame),  and (ii) the flux tube forming  the uniform constant external  magnetic field rests in the rotation frame.
Two limiting cases were studied analytically, when $R_s\simeq R$, and when $R_s\ll \xi_{>}\ll R$, where $\xi_{>}$ is the typical length scale, at which the vortex field changes in this region, see  Eq. (\ref{chiout}). For $R_s=R$ in case when the flux tube rests in the resting frame, at $A_0=0$ we recovered  results of \cite{Zahed,Guo} and  results \cite{Voskresensky:2023znr,Voskresensky:2024ivv,Voskresensky:2024vfx} for $eA_0=-V_0\leq 0$.

We demonstrated  for $\lambda =0$ that in the case (i) the ground state level lies deeper than in the  case (ii), see    (\ref{dispH}) and (\ref{dispHii}). This implies that if the solenoid of the radius $R_s=R$ was initially rotated with the cylinder of approximately the same radius, after a while it would be energetically profitable to stop the former. Next, we solved the problem for a narrow solenoid provided  $ \xi_{>}\gg R_s$. In this case  results are approximately the same (exactly the same in  limit $R_s\to 0$) independently if the flux tube rests in the resting or rotation frames.


 In Sec. \ref{acceleration} we considered the spinless bosons under action of  acceleration or deceleration  besides the  rotation. Specific example is studied explicitly when   the local acceleration/deceleration has the form  $w(r)=G/r$, for $G=const$. We demonstrated that even in case  when external electromagnetic field is switched off, $A_\mu=0$, there is a critical value of the rotation frequency $\Omega_c\approx (1+G)/R$ (for $l\gg mR\gg 1$) such that for $\Omega>\Omega_c$ for any deceleration, $G<0$, the rotating system  becomes unstable in respect to production of spinless bosons. With taking into account $\lambda |\phi|^4$ self-interaction and electromagnetic interaction the stability is recovered by formation of the spinless-vortex-field condensate. Switching on the electromagnetic field acts in favor of the formation of the  condensate vortex field.
In heavy-ion collisions at an initial stage the system undergoes  violent deceleration. Then the formed quasi-equilibrium nuclear  fireball may undergo first acceleration and then deceleration. Moreover in non-central heavy-ion collisions there appears rapid rotation and a strong magnetic field. Our estimates demonstrated that  these effects  may  stimulate possibility of creation of  a charged pion vortex field in heavy-ion collisions.

In the given work the electromagnetic field was considered as the external field. We may hope that this approximation may hold at least in case of a rather weak condensate field when the redistribution of the charge due to occurrence of the boson condensate  does not yet significantly modify the profile of the external field.

We thank H. Grigorian, E. E. Kolomeitsev and I. G. Pirozhenko  for fruitful discussions.

\appendix
\section{Interval in arbitrary moving frame}\label{Ap}
The interval and space-time metric in the accelerated,
rotating frame  with Cartesian spatial coordinates $x^i$
and time $t$ are  as follows \cite{Nelson1987,Voytik}
\begin{eqnarray} &(ds)^2=[(1+\vec{w}\vec{r}_3)^2-(\vec{\Omega}\times \vec{r}_3)^2)](dt)^2 -2(\vec{\Omega}\times \vec{r}_3)d\vec{r}_3 dt  - (dr_3)^2\,,\label{intervalaccrot}
  \end{eqnarray}
\begin{eqnarray}
g^{\mu\nu}=\frac{1}{(1+\vec{w}\vec{r}_3)^2}
  \begin{pmatrix}  1 & y\Omega & -x\Omega&0 \\
   y\Omega & -(1+\vec{w}\vec{r}_3)^2+y^2\Omega^2 & -xy\Omega^2&0 \\
   -x\Omega & -xy\Omega^2 & -(1+\vec{w}\vec{r}_3)^2+x^2\Omega^2&0 \\
   0 & 0 & 0&0 \\
\end{pmatrix}\,,\nonumber
\end{eqnarray}
\begin{eqnarray}
g_{\mu\nu}=
  \begin{pmatrix}  (1+\vec{w}\vec{r}_3)^2-r^2\Omega^2 & y\Omega & -x\Omega&0 \\
   y\Omega & -1 & 0&0 \\
   -x\Omega & 0 & -1&0 \\
   0 & 0 & 0&-1 \\
\end{pmatrix}\,,\label{gmunuac}
\end{eqnarray}
\begin{eqnarray}{\Box}=
\frac{\partial_\mu (\sqrt{-g}g^{\mu\nu}\partial_\nu)}{\sqrt{-g}}\,,\quad \mbox{det}g^{\mu\nu}=g=-(1+\vec{w}\vec{r}_3)^2\,,
\end{eqnarray}
and in these notations the Klein-Gordon equation renders
\begin{eqnarray}{\Box}\phi =m^2 \phi\,.
\end{eqnarray}
Here  $\vec{w}$ is (in general case time-dependent) acceleration
of the  reference frame  relative to the laboratory
(resting) frame ``$R$'', $\vec{r}_3$ is the position vector locating a
spatial point with respect to the origin of the reference frame, and $\vec{\Omega}$ is (in general case time-dependent) angular velocity of the moving frame  with respect to
the resting frame.  In case of our interest  the  moving frame is associated with the motion $\vec{\Omega}\parallel z$, $\Omega =const$ and we assume that $w={w}(r)$. We will conjecture that transformation (\ref{intervalaccrot}), (\ref{gmunuac}) holds in this case at least for a smooth $r$ dependence. The causality condition requires fulfilment of the inequality $g_{00}>0$. For $\Omega =0$ it requires that
 $wr>-1$.

\end{document}